\documentclass[aps,prl,twocolumn,longbibliography,floatfix]{revtex4-2}
\usepackage{amsmath}
\usepackage{amsfonts}
\usepackage{amssymb}
\usepackage{mathtools}
\usepackage[colorlinks=true,citecolor=blue,linkcolor=blue,urlcolor=blue]{hyperref}
\usepackage{bm}
\usepackage{times}
\usepackage{cases}
\usepackage{wasysym}
\usepackage[version=4]{mhchem}
\usepackage{graphicx}
\usepackage{subfigure}
\usepackage{comment}

 \graphicspath{ {./images/} }
 \usepackage{physics}
 
\begin{document}
\title{Upper branch thermal Hall effect in quantum paramagnets}

\author{Bowen Ma$^{2,3}$}
\author{Z. D. Wang$^{2}$} 
\author{Gang Chen$^{1,2,3,}$}
\email{gangchen.physics@gmail.com}
\affiliation{$^{1}$International Center for Quantum Materials, 
School of Physics, Peking University, Beijing 100871, China}
\affiliation{$^{2}$Department of Physics and HK Institute of Quantum Science \& Technology,  
The University of Hong Kong, Pokfulam Road, Hong Kong, China}
\affiliation{$^{3}$The University of Hong Kong Shenzhen Institute of Research and Innovation, 
Shenzhen 518057, China}

\date{\today}
    
\begin{abstract}
Inspired by the persistent thermal Hall effects at finite temperatures in various quantum 
paramagnets, we explore the origin of the thermal Hall effects from the perspective 
of the upper branch parts by invoking the dispersive and twisted crystal field excitations. 
It is shown that, the upper branches of the local energy levels could hybridize and form 
the dispersive bands. The observation is that, upon the time-reversal symmetry breaking 
by the magnetic fields, these upper branch bands could acquire a Berry curvature distribution 
and contribute to the thermal Hall effect in the paramagnetic regime. As a proof of principle, 
we consider the setting on the kagom\'e lattice with one ground state singlet and an excited 
doublet, and show this is indeed possible. We expect this effect to be universal and 
has no strong connection with the underlying lattice. Although the thermal Hall signal
can be contributed from other sources such as phonons and their scattering in the actual materials, 
we discuss the application to the Mott systems with the large local Hilbert spaces. 
\end{abstract}

\maketitle

Recently thermal Hall transports are widely used to explore the properties of the 
elementary excitations in correlated quantum materials. 
In the Mott insulating systems where the relevant excitations are charge neutral, 
the thermal Hall effect plays an important role in deciphering the Berry curvature 
properties of the excitations~\cite{zhang2023thermal}. 
For spin liquids, the half-quantized thermal Hall conductivity is one smoking-gun result 
for the gapped Kitaev spin liquid with the chiral Majorana edge mode~\cite{kitaev2006anyons}, 
and might have been observed in $\alpha$-RuCl$_3$~\cite{kasahara2018majorana,yokoi2021half}. 
The thermal Hall effects could reflect the intrinsic matter-gauge coupling and 
the Berry curvature properties of the emergent exotic quasiparticles in different spin liquids. 
As a probe of the magnetic excitations, the thermal Hall effect is found to be useful 
in more conventional magnets. The magnon thermal Hall 
effects~\cite{onose2010observation,katsura2010theory,laurell2018magnon,neumann2022thermal,mcclarty2022topological} 
were widely studied in many ordered magnets. In a class of magnets known as 
``dimerized magnets'' where the ground state is approximately given as the product 
of the spin-singlet dimers on the bonds with stronger exchanges, 
the spin-triplet excitations, known as ``triplons'', can propagate via the inter-dimer couplings 
and form the triplon bands. These triplon bands can acquire the non-trivial Berry curvatures 
and even finite Chern numbers once the anisotropic interaction such as the Dzyaloshinskii-Moriya 
interaction is introduced~\cite{romhanyi2015hall,mcclarty2017topological}. 
This leads to interesting behaviors in the triplon thermal Hall effect.

The thermal Hall conductivity has been measured in several visible quantum magnets 
such as Tb$_2$Ti$_2$O$_7$~\cite{hirschberger2015large}, Pr$_2$Zr$_2$O$_7$~\cite{chu2023low}, 
Pr$_2$Ir$_2$O$_7$~\cite{uehara2022phonon} and Na$_2$Co$_2$TeO$_6$~\cite{yang2022significant,takeda2022planar}. 
One common feature of these quantum magnets is that, due to the combination 
of the crystal electric field (CEF) and the spin-orbit coupling (SOC),   
the magnetic ions have a relatively large local physical Hilbert space  
with a series of local energy levels~\cite{gardner2010magnetic}. 
We take the well-known compound Tb$_2$Ti$_2$O$_7$ 
for example~\cite{gardner1999cooperative,PhysRevLett.98.157204,hirschberger2015large}. 
Via the SOC, the Tb$^{3+}$ ion has a ${J = 6}$ local moment. 
The 13-fold degeneracy is further split by the CEF into multiple singlets 
and doublets. As the thermal Hall transport in Tb$_2$Ti$_2$O$_7$ was 
measured up to 142 K and 10 T~\cite{hirschberger2015large}, 
at this temperature scale, the second excited doublet (at 1.41 meV) has already been 
thermally activated. The CEF multiplets could then make a significant impact 
on the physics of those quantum magnets~\cite{voleti2023impact}. In addition, 
the CEF levels at 10-20 meV would be thermally populated. 
The 10 T magnetic field reorganizes these CEF states and splits the doublets.
For the activated CEF states, the 10 T field could create a Zeeman splitting of about 20-40 K. 
The previous work that studies the monopole thermal Hall effect~\cite{zhang2020topological}  
from the ground state doublets in the quantum ice regime~\cite{PhysRevLett.98.157204} 
certainly cannot be extended to such high temperatures and large magnetic field regimes. 
Thus, these two ingredients, i.e. the thermal activation and the field 
splitting/hybridization of the CEF states, indicate that, 
one should seriously consider the involvement of these excited  
CEF states in thermal transports. Similar physics should  
generally occur in other quantum magnets with a large local Hilbert space. 
This aspect is quite different from the cuprate system where the local Hilbert space 
for a large range of energy scale is a spin-1/2 local moment from the $e_g$ electrons 
and a large thermal Hall signal was observed~\cite{grissonnanche2019giant,boulanger2020thermal}. 
Remarkably, if one views these CEF excitations as the generalized ``triplons'' 
with respect to the CEF ground state, this view bridges 
this series of quantum magnets with the dimerized magnets. 
One immediate outcome is that, these generalized ``triplons'', 
similar to the topological excitations in excitonic magnets~\cite{anisimov2019nontrivial}, 
should in principle possess the Berry curvatures in the magnetic field 
and contribute to the thermal Hall conductivity at the relevant temperature regime. 
In Ref.~\cite{liu2019upper}, the upper branch magnetism 
from the excited CEF states was understood when the CEF gap is comparable 
to the exchange interaction between the CEF states of neighboring sites. 
 Thus, the thermal Hall effect from the generalized triplons is dubbed 
``upper branch thermal Hall effect''.

\begin{figure}
\subfigure[]{\label{fig:Kagome}\includegraphics[width=0.34\textwidth]{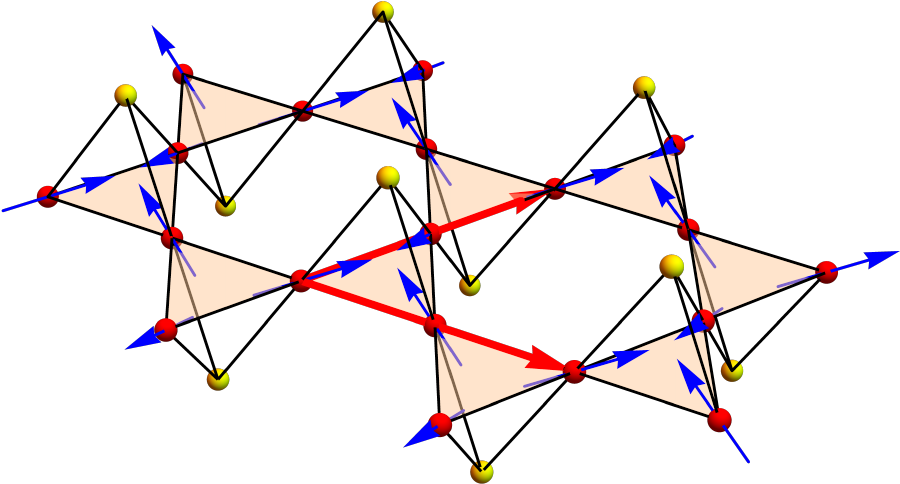}}\subfigure[]
{\label{fig:Tripod}
\includegraphics[width=0.14\textwidth]{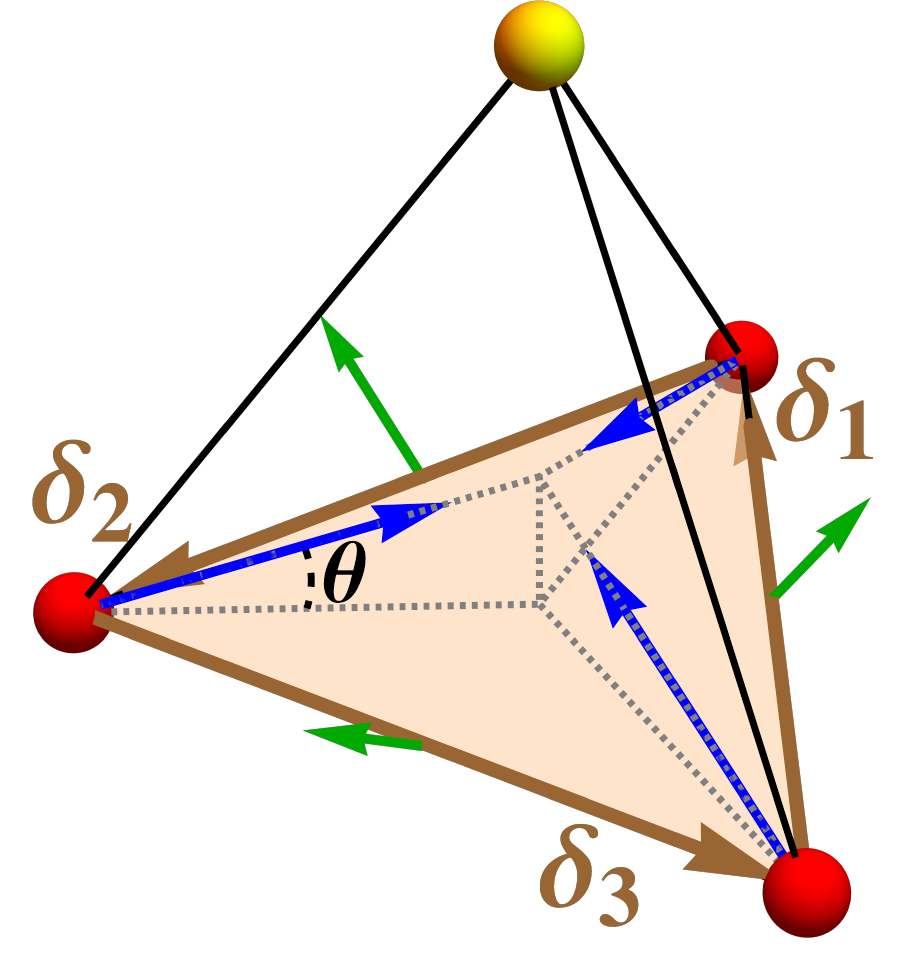}}
    \caption{
    (a) The tripod kagom\'e lattice with the effective spins on the red sites. 
    The blue arrows denote the local Ising $z$ axes. 
    The red vectors are the basis vectors ${\mathbf{a}_1=a(1,0)}$ 
    and ${\mathbf{a}_2=a(1/2,\sqrt{3}/2)}$. 
    (b) The tripod unit cell. The Ising $z$ axes have a canting angle $\theta$ with the kagom\'e plane. 
    The green arrow normal to each bond shows the DM vectors with in-plane component $D_p$ 
    and out-of-plane component $D_z$. The neighboring bonds are ${\boldsymbol{\delta}_1=a(-1/4,\sqrt{3}/4)}$, 
    $\boldsymbol{\delta}_2=a(-1/4,-\sqrt{3}/4)$, and $\boldsymbol{\delta}_3=a(1/2,0)$.}
    \label{fig:Lattice}
\end{figure}

Since the upper branch thermal Hall effect arises from the CEF states and their interactions that
depend on the CEF wavefunctions and lattice symmetries, the underlying lattice is needed 
but does not play a significant role in the physics. Thus, we simply consider the setting  
on a kagom\'{e} lattice as shown in Fig.~\ref{fig:Lattice}. 
This can either be established by applying the [111] magnetic field 
on the pyrochlore magnets, or naturally occurs 
in the tripod kagom\'{e} magnets~\cite{dun2016magnetic,dun2020quantum}. 
Since we are mostly concerned about the excited CEF states, to simplify the problem, 
we assume the CEF ground state is a singlet and the excited CEF states form a doublet. 
The three states of the local moment are then described by an effective spin ${S=1}$ with
an onsite anisotropic term $\sum_i \eta(\hat{\mathbf{z}}_i\cdot\mathbf{S}_i)^2$ and ${\eta >0}$,
where $\hat{\mathbf{z}}_i$ is defined along the local coordinate system for each sublattice.  
The spin-1 moment differs from the pseudospin-1/2 moment that is often used to describe Kramers 
or non-Kramers doublets. In our design, the lower singlet corresponds to ${S_i^{z_i} =0}$,
and the upper doublet correspond to ${S_i^{z_i} = \pm 1}$. If the lower singlet and the upper 
doublet are connected by the ladder operator of the original $J$ operators, 
all the components of ${\mathbf S}_i$ are odd under time reversal.

For the spin interaction, we consider an exchange model that is quadratic 
in the effective spin-1 components. The effective spin Hamiltonian is written 
as
\begin{eqnarray}
    H&=& \sum_{\langle ij\rangle}\big[J \mathbf{S}_i\cdot\mathbf{S}_j
    +\mathbf{D}_{ij}\cdot (\mathbf{S}_i\times\mathbf{S}_j ) \big]
+\sum_i\eta({S}_i^{z_i})^2    
    \nonumber \\
    && - \sum_i B(\hat{\mathbf{z}}\cdot\hat{\mathbf{z}}_i) {S}_i^{z_i} ,
    \label{H}
\end{eqnarray}
where $\mathbf{D}_{ij}$ is the Dzyaloshinskii-Moriya (DM) vector~\cite{moriya1960anisotropic,dzyaloshinsky1958thermodynamic}    
for the bond $ij$ with both in-plane component $D_p$ and out-of-plane component $D_z$ in general~\cite{sm}. 
Although the Zeeman coupling could involve all the spin components, 
only the Zeeman coupling to the local $z$ component is considered for simplicity.     
Moreover, more complicated spin interactions such as the pseudo-dipolar interaction  
and higher-order spin interactions could be present. This kind of interaction between the upper 
doublets of neighboring sites have the form of a four-spin interaction~\cite{liu2019upper}.

\begin{figure}
\includegraphics[width=7.5cm]{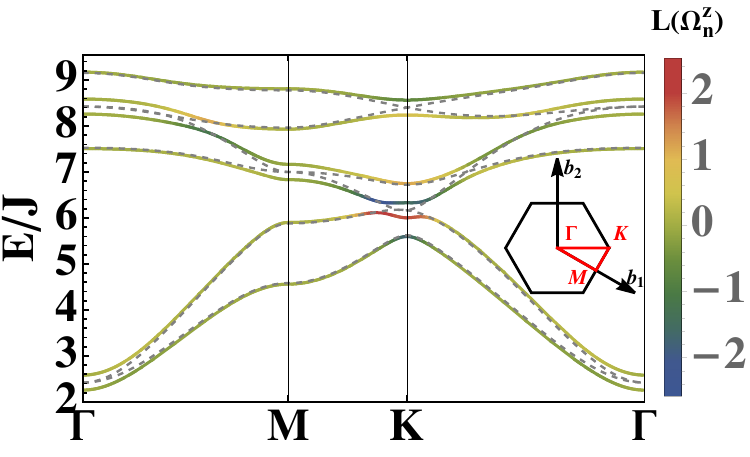}
\caption{The band dispersion of the doublet excitations from the linear flavor-wave theory. 
We set ${\sin\theta=1/3}$, ${\eta/J=7.0}$, $\sqrt{3}D_p=\sqrt{3/2}D_z=D=0.9J$, 
and ${B/J=0.5\ (0)}$ for the solid (dashed) lines. 
The color of the solid line shows the non-zero Berry curvature~\cite{sm} in the log scale $\text{L}(\Omega_n^z)=\text{sgn}(\Omega_n^z)\text{ln}(1+|\Omega_n^z|)$. 
The inset shows the hexagonal Brillouin zone.
 }
\label{fig:Dispersion} 
\end{figure}


In the strong anisotropic limit with ${\eta>0}$, the ground state is a simple 
quantum paramagnet with ${S_i^{z_i} =0}$. With the exchange interaction, 
the many-body ground state depletes a bit from ${S_i^{z_i} =0}$, which is analogous 
to the depletion of superfluid weight in the Bose-Einstein condensation of interacting 
bosons. The excited doublets form the dispersive bands. 
The picture does not alter much in the 
presence of external magnetic fields. As the ground state is paramagnetic without ordering, 
the usual Holstein-Primakoff boson representation is not suitable to describe the excitations. 
Instead, we regard ${{S}_i^{z_i}=0,\pm 1}$ 
as three different flavors~\cite{joshi1999elementary,li19984} in the spirit of the SU(3) flavors, 
and invoke a flavor representation of $\mathbf{S}_i$ as~\cite{li2018competing,sm},
\begin{align}
    \left\{\begin{array}{lll}
     S_i^{z_i}&\equiv\hat{\mathbf{z}}_i\cdot\mathbf{S}_i=b^\dagger_i b_i^{}-\bar{b}^\dagger_i \bar{b}_i^{}, \vspace{1mm} \\
     S_i^-&\equiv(\hat{\mathbf{x}}_i-i\hat{\mathbf{y}}_i)\cdot\mathbf{S}_i=\sqrt{2}(\bar{b}^\dagger_i + b_i^{} ),\vspace{1mm}  \\
     S_i^+&\equiv(\hat{\mathbf{x}}_i+i\hat{\mathbf{y}}_i)\cdot\mathbf{S}_i=\sqrt{2}(\bar{b}_i^{} +b^\dagger_i),
    \end{array}
    \right.
\end{align}
where $(\hat{\mathbf{x}}_i,\hat{\mathbf{y}}_i,\hat{\mathbf{z}}_i)$ spans a local coordinate for site $i$. 
For the quantum paramagnet here, the boson operators $b^\dagger_i$ and $\bar{b}^\dagger_i$ 
($b_i$ and $\bar{b}_i$) create (annihilate) a state with a ``magnetic flavor'' 
$S_i^{z_i} =+1$ and $-1$, respectively.
Due to the noncollinearity of the Ising axes of the three sublattices, 
one needs to rotate the spin operators in Eq.~\eqref{H} into the local coordinate 
for different sites~\cite{del2004quantum,ma2020longitudinal}, 
and this generates the pairing of the flavor bosons. 
Therefore, after the Fourier transform, the Hamiltonian Eq.~\eqref{H} 
needs to be written in a Bogoliubov-de Gennes (BdG) form~\cite{sm} as 
$H=\frac{1}{2}\sum_\mathbf{k}\Psi_\mathbf{k}^\dagger H_\mathbf{k}^{} \Psi_\mathbf{k}^{}$ 
with
\begin{align}
    H_\mathbf{k}=\begin{pmatrix}
        A_\mathbf{k} & B_\mathbf{k}\\
        B^*_\mathbf{-k} & A^*_\mathbf{-k}
    \end{pmatrix},
\end{align}
and
\begin{equation}
\Psi_\mathbf{k}\! =\! \left(b_{1\mathbf{k}}^{}, \bar{b}_{1\mathbf{k}}^{},...,b_{3\mathbf{k}}^{}, \bar{b}_{3\mathbf{k}}^{}, 
b^\dagger_{1,-\mathbf{k}},\bar{b}^\dagger_{1,-\mathbf{k}},..., b^\dagger_{3,-\mathbf{k}},\bar{b}^\dagger_{3,-\mathbf{k}}\right)^T. 
\end{equation}
The dispersion of the flavor-wave excitations can be determined as the positive eigenvalues of 
$\Sigma_z H_\mathbf{k}$~\cite{del2004quantum,ma2022antiferromagnetic}, 
where ${\Sigma_z=\sigma_z \otimes I_{6}}$ 
with $\sigma_z$ the Pauli matrix and $I_{n}$ the ${n\times n}$ identity matrix. In Fig.~\ref{fig:Dispersion}, 
we depict the representative dispersions in the quantum paramagnetic phase.
When ${B\neq 0}$, the bands are separated from each other. 
If the band bottom touches zero energy, the bosons begin to condensate~\cite{sm}, 
and the system develops a corresponding magnetic order~\cite{li2018competing}. 
To avoid that, we work in the regime with ${B<\eta}$ such that the quantum 
paramagnet remains stable throughout.


In magnetically ordered systems,  
the DM interaction or/and the noncollinear spin configuration could give   
rise to topological magnons with non-zero Chern numbers~\cite{mook2014edge,kim2016realization,owerre2016first,laurell2018magnon}. 
With the DM interactions and/or the noncollinear Ising axes in the current model, 
the Berry curvature of the flavor-wave excitation in the quantum paramagnet 
is also expected to be non-zero. In the case of the bosonic BdG Hamiltonian, 
the wavefunction of the $n$-th band $|\psi_{n\mathbf{k}}\rangle$ is determined by the eigen-equation 
$E_{n\mathbf{k}}|\psi_{n\mathbf{k}}\rangle=\Sigma_z H_{\mathbf{k}}|\psi_{n\mathbf{k}}\rangle$. 
The corresponding Berry connection $\boldsymbol{\mathcal{A}}_{n\mathbf{k}}$ 
and Berry curvature $\boldsymbol{\Omega}_{n\mathbf{k}}$ is then defined as~\cite{mcclarty2022topological,sm},
\begin{align}
    \boldsymbol{\mathcal{A}}_{n\mathbf{k}}=i\langle \psi_{n\mathbf{k}}|\Sigma_z\boldsymbol{\nabla}_\mathbf{k}|\psi_{n\mathbf{k}}\rangle,\text{ and }\boldsymbol{\Omega}_{n\mathbf{k}}=\boldsymbol{\nabla}_{\mathbf{k}}\times\boldsymbol{\mathcal{A}}_{n\mathbf{k}}.
\end{align}
In two-dimensional systems, the first Chern number can then be calculated 
by integrating the $z$-component of $\Omega_{n\mathbf{k}}$ over the Brillouin zone as
\begin{align}
    C_n=\frac{1}{2\pi}\sum_{\mathbf{k}}\Omega_{n\mathbf{k}}^z.
\end{align}

\begin{figure}
\includegraphics[width=0.48\textwidth]{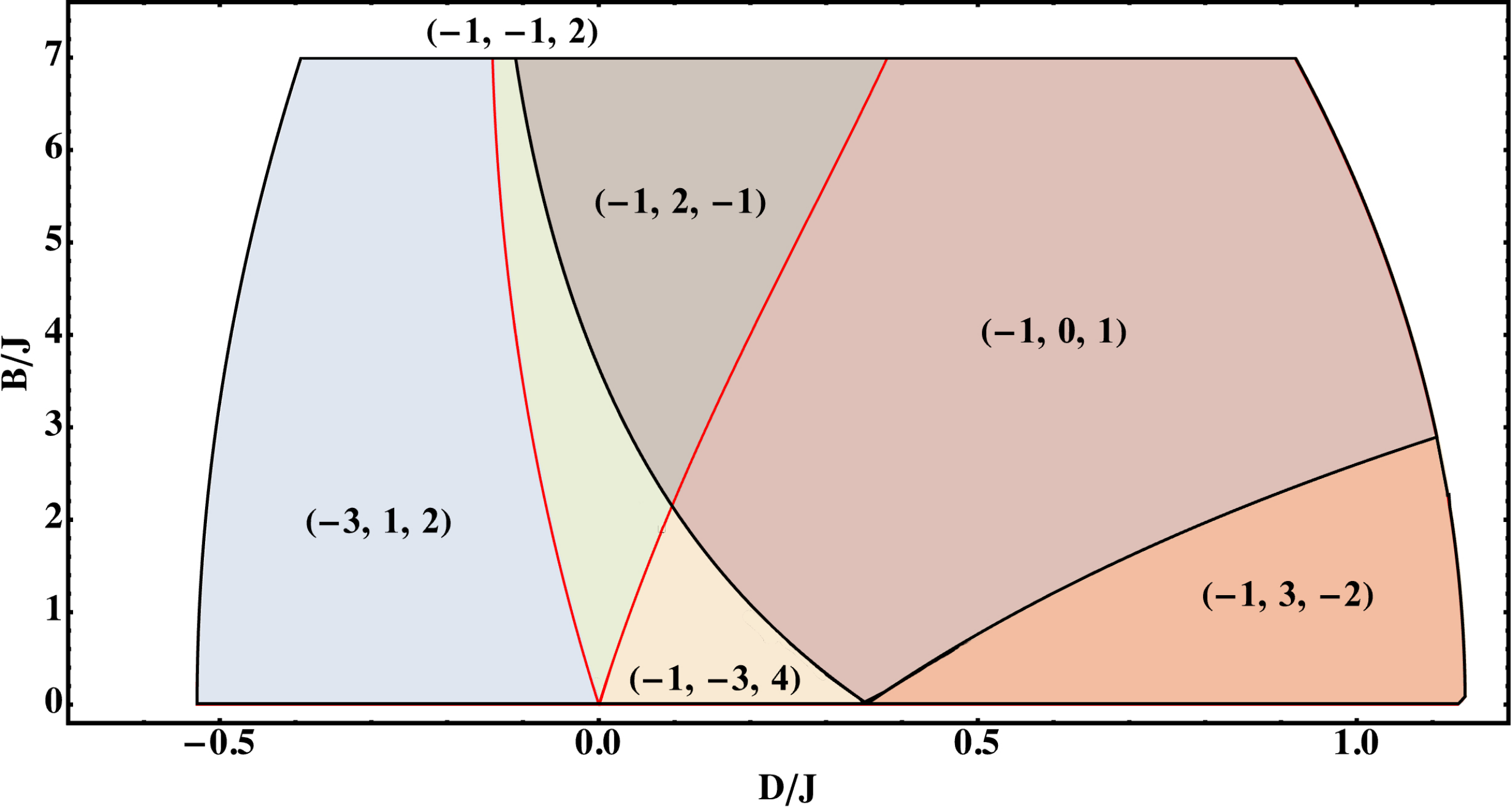}
\caption{Diagram of Chern numbers distributions for the lower three bands. 
The Chern numbers are listed from bottom to top. We set 
${\sin\theta=1/3}$, ${\eta/J=7.0}$, $\sqrt{3}D_p=\sqrt{3/2}D_z=D$, and ${B<\eta}$. 
The red solid and black thick lines denote the band touching
and the Chern number change at high-symmetry points $\boldsymbol{\Gamma}$ and 
$\mathbf{M}$, respectively. 
The outer boundary of the quantum paramagnet is determined 
when the band bottom touches zero energy at $\boldsymbol{\Gamma}$.}
\label{fig:CN} 
\end{figure}

The general analysis of the band topology with an arbitrary choice of 
parameters is unnecessary for our purpose. Without loss of much generality, 
we consider a simple case where the Ising axes $\hat{\mathbf{z}}_i$ 
are all perpendicular to the kagom\'e plane, and DM vectors only have 
out-of-plane component as $\mathbf{D}_{12}=\mathbf{D}_{23}=\mathbf{D}_{31}=D_z\hat{\mathbf{z}}$. 
With this simplification, if we perform a basis transformation that   
${u_{m\mathbf{k}}=(b_{m\mathbf{k}}^{}+\bar{b}_{m,-\mathbf{k}}^\dagger)/\sqrt{2}}$ 
and ${p_{m\mathbf{k}}=i(\bar{b}_{m,-\mathbf{k}}^\dagger-b_{m\mathbf{k}})/\sqrt{2}}$, 
the Hamiltonian can then be written as 
\begin{align}
    H&=\frac{1}{2(2\eta)^{-1}}\sum_\mathbf{k}
    (\mathbf{p}_\mathbf{k}^\dagger    -i{B}{\eta}^{-1} \mathbf{u}^\dagger_\mathbf{k})
    (\mathbf{p}_\mathbf{k}    + i {B}{\eta}^{-1} \mathbf{u}_\mathbf{k})
    \nonumber\\
    &+\frac{1}{2}\sum_\mathbf{k}\mathbf{u}_\mathbf{k}^\dagger(4M_\mathbf{k}  -  {2B^2}{\eta}^{-1})\mathbf{u}_\mathbf{k}
    \label{phonon}
\end{align}
with
\begin{align}
    M_{\mathbf{k}}&=\frac{\eta}{2}I_3+\begin{pmatrix}
        0 & 2\tilde{J}\cos\mathbf{k}_3 & 2\tilde{J}^*\cos\mathbf{k}_2\\
        2\tilde{J}^*\cos\mathbf{k}_3 & 0 & 2\tilde{J}\cos\mathbf{k}_1\\
        2\tilde{J}\cos\mathbf{k}_2 & 2\tilde{J}^*\cos\mathbf{k}_1 & 0
    \end{pmatrix}
    \label{M}
\end{align}
where ${\tilde{J}=J+iD_z=|\tilde{J}|e^{-i\phi/3}}, {\mathbf{k}_m=\mathbf{k}\cdot\boldsymbol{\delta}_m}$,$ 
\mathbf{u}_\mathbf{k}=(u_{1\mathbf{k}}^{},\\u_{2\mathbf{k}}^{},u_{3\mathbf{k}}^{})$, 
${\mathbf{p}_\mathbf{k}=(p_{1\mathbf{k}}^{}, p_{2\mathbf{k}}^{}, p_{3\mathbf{k}}^{} )}$ 
and $[u^\dagger_{m\mathbf{k}},p_{m'\mathbf{k}'}^{}]=i\delta_{mm'}\delta_{\mathbf{k}\mathbf{k}'}$. 
The Hamiltonian Eq.~\eqref{phonon} is an analog of a phononic system on the kagom\'e lattice 
with a mass $(2\eta)^{-1}$ and a dynamical matrix $4M_\mathbf{k}-{2B^2}{\eta}^{-1}$. 
It can be shown~\cite{sm} that the wavefunction of this ``phononic'' Hamiltonian remains unchanged 
when ${B=0}$ and the topological properties are actually determined by $M_\mathbf{k}$. 
Interestingly, $M_\mathbf{k}$ is topologically equivalent to the chiral spin Hamiltonian~\cite{ohgushi2000spin} or topological magnonic Hamiltonian~\cite{katsura2010theory} on kagom\'e lattice, where the inequivalence between the honeycomb plaquette and the triangular plaquette leads to a non-zero $\phi$-flux. Therefore, there is a SU(3)$\oplus$SU(3) band topology where the Chern numbers of the three bands with flavor $\pm 1$ are determined by a SU(3) structure~\cite{barnett20123} as $\left(\mp\text{sgn}(\sin\phi),0,\pm\text{sgn}(\sin\phi)\right)$ from bottom to top.

In a more general situation with non-collinear Ising axes, we choose ${\eta=7J}$ 
and assume the system further respects a $O_h$ point group symmetry 
that is inherited from the parent pyrochlore lattice so that $D_z=\sqrt{2}D_p=\sqrt{2/3} D$. 
After numerically computing the band Chern number in the discretized 
momentum space~\cite{fukui2005chern}, we obtain a topological phase diagram for the lower three bands shown in Fig.~\ref{fig:CN}. The full diagram for all six bands can be found in the supplement~\cite{sm}. We find that, due to the mixing of the two flavors in the global coordinate, the SU(3)$\oplus$SU(3) topology is enriched with varying intrinsic couplings as well as the external magnetic field.

From the analytical calculation and numerical study above, we can see that non-trivial Berry physics of the excited doublets in our model originates from the non-cancellation of the flux in the kagom\'e lattice, and thus we believe that similar non-trivial topology for even more general multiplet excitations will also occur in various lattices with inequivalent plaquettes such as honeycomb~\cite{ganesh2011quantum}, checker-board~\cite{moessner2004planar,sadrzadeh2019quantum} and bulk~\cite{li2018competing} or thin-filmed~\cite{hu2012topological} pyrochlore lattices.

\begin{figure}
{\includegraphics[width=6.4cm]{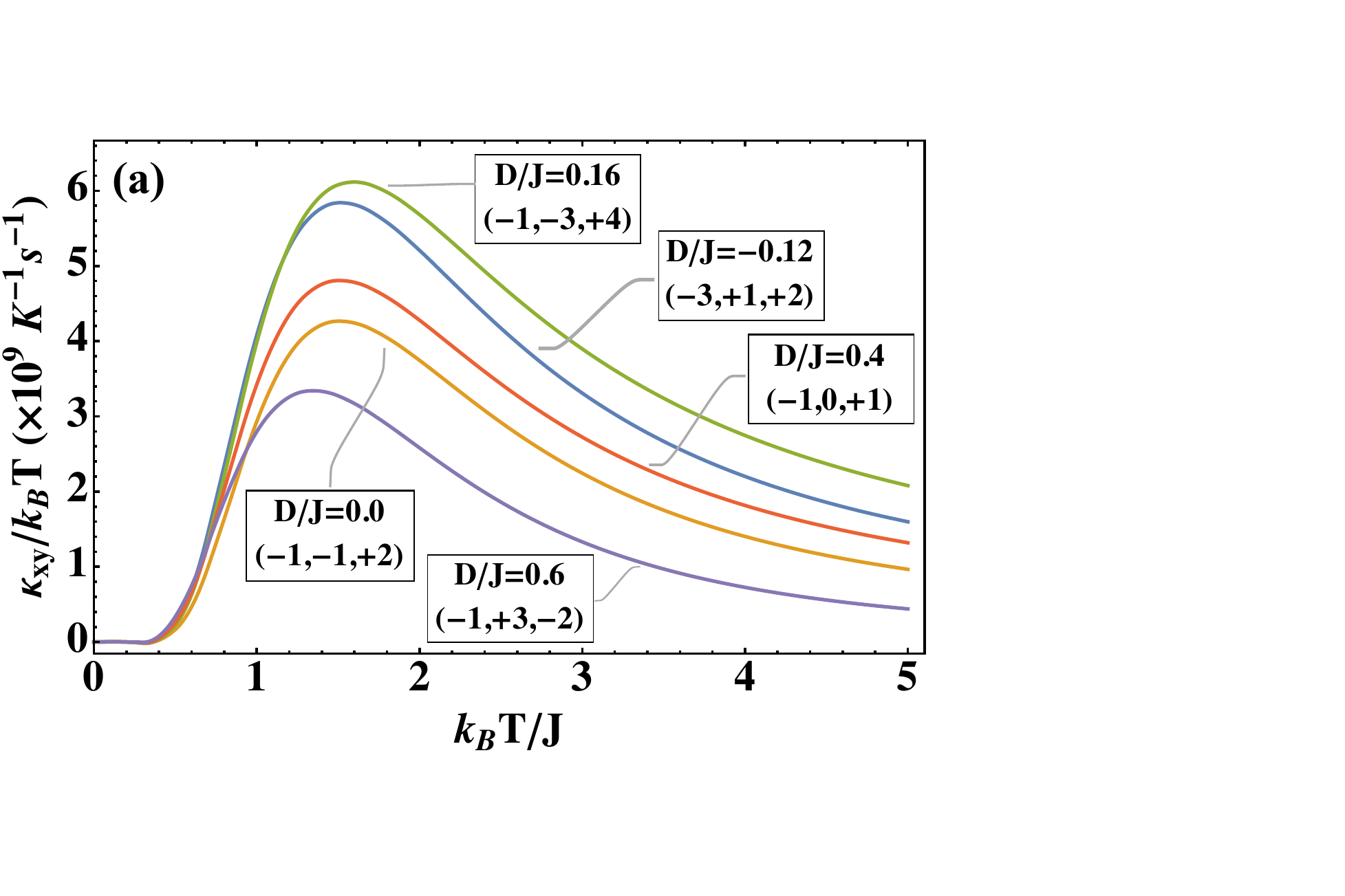}\label{fig:THC-DMI}}
{\includegraphics[width=7.26cm]{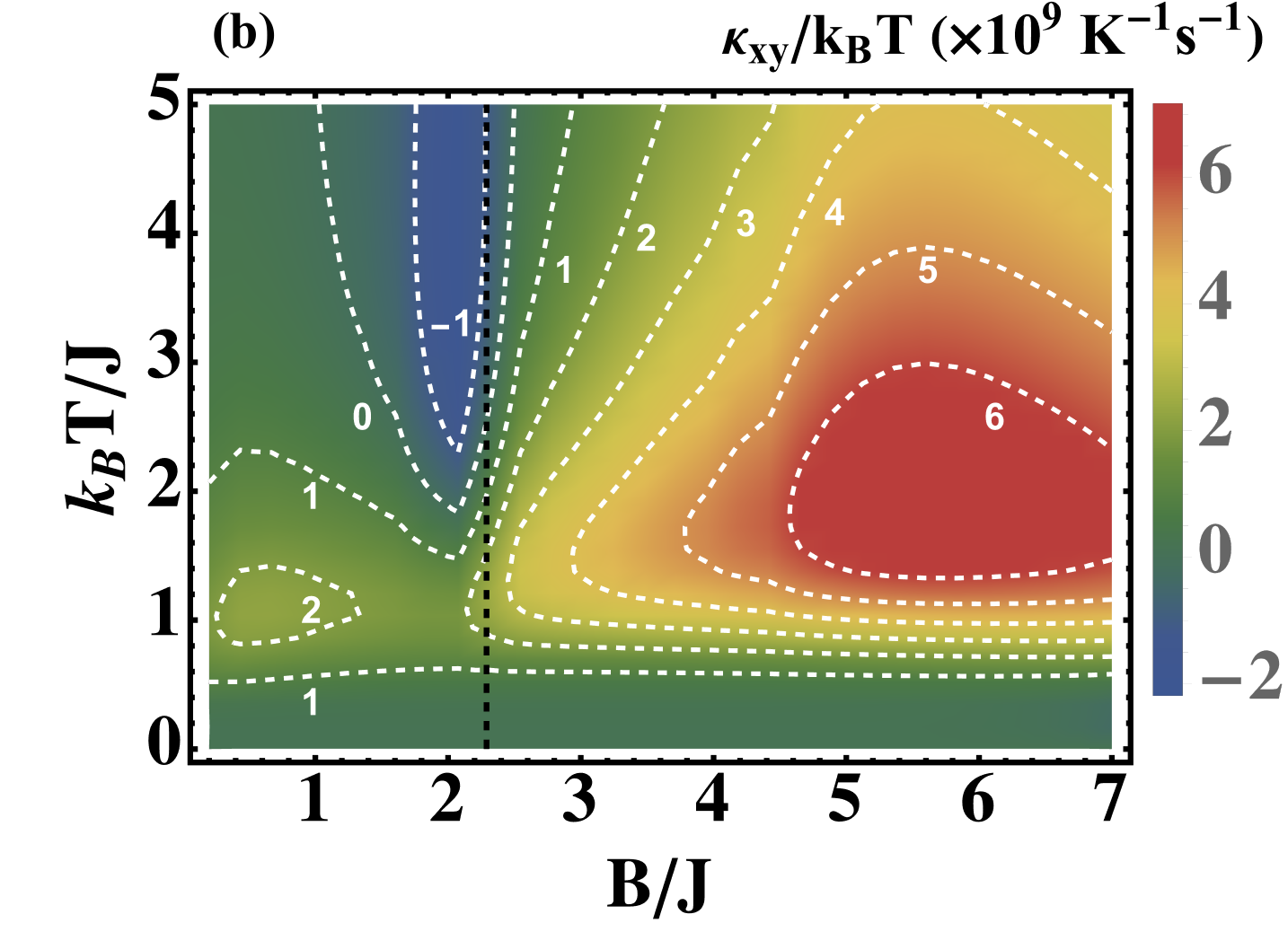}\label{fig:THC-Density}}
    \caption{The thermal Hall conductivity with varying parameters. 
    We set $\eta/J=7.0$, ${\sqrt{3}D_p=\sqrt{3/2}D_z=D}$, and (a) ${B/J=0.8}$ with different values of $D$ 
    and the corresponding Chern numbers of the lower three bands labeled in the plot; (b) ${D/J=0.9}$ 
    with the white numbers and dashed lines denoting the values of ${\kappa_{xy}}/{(k_B T)}$ (in unit of $10^9$ K$^{-1}\cdot$s$^{-1}$). 
    The black dashed line shows the critical field ${B_c/J\approx 2.3}$ 
    where the Chern numbers of the lower three bands change between $(-1,+3,-2)$ and $(-1,0,+1)$, 
    resulting in a sign change of the thermal conductivity.}
    \label{fig:THC}
\end{figure}

Semiclassically, with the finite Berry curvature, 
the wave-packet of the excitations will experience an anomalous velocity from $\boldsymbol{\Omega}$  
as~\cite{xiao2010berry,cheng2016spin},
\begin{align}
    \dot{\mathbf{r}}_n=\frac{1}{\hbar}\frac{\partial E_{n\mathbf{k}}}{\partial \mathbf{k}}-\dot{\mathbf{k}}\times\boldsymbol{\Omega}_{n\mathbf{k}},
\end{align}
where $\mathbf{r}_n$ is the packet center of the $n$-th wavefunction.
If a longitudinal temperature gradient $\nabla_yT$ is applied across the material, 
the transverse motion of the excitations from the anomalous velocity term will lead to some Hall-like transport signals. In the case of thermal Hall effects, the excitations carrying different energies will experience different anomalous velocities, and thus lead to a transverse temperature difference. From the theoretical side, the associated thermal Hall conductivity $\kappa_{xy}$ 
of bosonic excitations can be derived from linear response theory as~\cite{matsumoto2011rotational,matsumoto2011theoretical},
\begin{align}
    \kappa_{xy}=-\frac{k_B^2T}{\hbar V}\sum_{n,\mathbf{k}}\left[c_2(g(E_{n\mathbf{k}}))-\frac{\pi^2}{3}\right]\Omega_{n\mathbf{k}}^z,
    \label{THC}
\end{align}
where $c_2(x)=(1+x)\ln^2(1+1/x)-\ln^2x-2\text{Li}_2(-x)$, Li$_2(x)$  
is the polylogarithm function, $T$ is the average temperature, 
$V$ is the volume of the material, and $g(x)=(\text{exp}(x/k_B T)-1)^{-1}$ 
is the Bose-Einstein distribution. 
In Fig.~\ref{fig:THC}(a), we show the dependence of $\kappa_{xy}/k_BT$ 
on the temperature $T$ with different DM interactions $D$. 
As we can infer from Eq.~\eqref{THC}, because of the distribution function, 
the Berry curvature from lower bands contributes more to the thermal Hall conductivity. 
Therefore, the thermal Hall conductivity is large when ${D/J=0.16}$, 
as the two lowest bands both have negative Chern numbers. 
Meanwhile, the thermal Hall conductivity is small in the case of ${D/J=0.16}$, 
where the second lowest band with a large positive Chern number $+3$ 
suppresses the contribution of negative Berry curvature from the lowest band. 
Besides, if the temperature is getting higher, the occupations of the excitations 
in all six bands become more equally populated, and thus $\kappa_{xy}$ 
goes closer to zero owing to the fact that $\sum_{n,\mathbf{k}}\Omega_{n\mathbf{k}}^z=0$.

In experiments, the DM interaction is usually not tunable, and we thus depict a density plot of $\kappa_{xy}/k_BT$ with respect to the magnetic field $B$ and the temperature $T$ in Fig.~\ref{fig:THC}(b). With the lattice constant 10 \AA\ of Tb$_2$Ti$_2$O$_7$ as an estimate for the interlayer distance $l$, $\kappa_{xy}/k_BT\sim 10^9$ K$^{-1}$s$^{-1}$ gives rise to a bulk thermal Hall signal $\kappa_{xy}/lT\sim 10^{-5}$ WK$^{-2}$m$^{-1}$, in the same order as experimental measurements~\cite{hirschberger2015large}. Due to the change of lower-band Berry curvature, with the magnetic field increasing, when the Chern numbers of the two lowest bands change from $(-1,+3)$ to $(-1,0)$, there is a sign flip of the thermal Hall conductivity around the critical field (denoted as the dotted black line). It should be pointed out that the phononic or extrinsic contribution to the sign change of thermal Hall effects in paramagnets is usually not tunable by the magnetic field or can be tuned limitedly accompanied by some magnetic phase transitions~\cite{chen2022large,li2023phonon}. We expect the observation of this sign change can indicate the presence of the upper branch thermal Hall effect, while a delicate experimental design may be needed to subtract the contribution from phonons as their effects are usually within the same order~\cite{uehara2022phonon}. As we stated in the previous section, though we obtain the above results from a specific model on the kagom\'e lattice, the tunable thermal Hall signal arising from the topology of the excited multiplets can generally occur in various lattices.

In this work, we have addressed the question whether applying a Zeeman field  
to Mott insulators with multiple local energy levels could generate the intrinsic   
thermal Hall effect solely from the magnetic excitations in the quantum paramagnetic 
phase at finite temperatures. In our simple modeling, we have only considered 
the lowest few CEF energy levels, which is sufficient to provide a positive answer. 
In reality, the candidate Mott insulators have many such CEF energy levels, 
and as the temperature increases, these CEF energy levels would be gradually 
thermally activated and contribute to thermal Hall transports. 
Thus, a comprehensive understanding of the thermal Hall signals
in the candidate materials requires the intrinsic components. 
We expect our results to be complementary to the recent efforts 
in the phonon thermal Hall effects with the non-Kramers-like doublet systems~\cite{guo2022resonant,guo2023phonon}. 
The previous analysis on Pr$_2$Ir$_2$O$_7$ has pointed out the resonant 
phonon-pseudospin scattering where the non-Kramers pseudospin arises 
from the ground state doublet of the Pr$^{3+}$ ion. 
The inclusion of the upper branch CEF states not only generates the intrinsic thermal Hall
sign as the upper branch thermal Hall effect, 
and may induce a cascade of resonant phonon scattering with the large local Hilbert space~\cite{chen2021mott}.

In conventional ordered magnets, the phonon-magnon hybridization~\cite{zhang2019thermal,go2019topological,zhang20203} was known to create Berry 
curvature distribution for the hybridized excitations, and can also lead to non-zero magneto-phonon chirality with
thermal Hall effects~\cite{ma2022antiferromagnetic,ma2023chiral}. This effect occurs even for the trivial magnon band structure that is absent of 
finite magnon Berry curvatures. 
For the exciton-like flavor-wave excitation in the quantum paramagnets,
similar phonon-exciton hybridization could occur. Here the flavor-wave excitation already 
develops Berry curvature distribution on its own. 
Thus, the hybridization could bring more interesting aspects to the dynamical properties of the whole system.

\emph{Acknowledgments.}---This work is supported by the National Science Foundation of China with Grant No.~92065203, the Ministry of Science and Technology of China with Grants No.~2021YFA1400300, and by the Research Grants Council of Hong Kong with Grant No.~C6009-20G and C7012-21G.

\bibliography{upperTHE}

\widetext
\clearpage
\begin{center}
\textbf{\large Supplementary Materials for ``Upper branch thermal Hall effect in quantum paramagnets''}
\end{center}
\begin{center}
    Bowen Ma$^{2,3}$, Z. D. Wang$^{2}$, and Gang Chen$^{1,2,3}$
\end{center}
\begin{center}
    \textit{\small $^1$International Center for Quantum Materials, School of Physics, Peking University, Beijing 100871, China\\
    $^2$Department of Physics and HK Institute of Quantum Science \& Technology,\\
The University of Hong Kong, Pokfulam Road, Hong Kong, China\\
    $^3$The University of Hong Kong Shenzhen Institute of Research and Innovation, Shenzhen 518057, China}
\end{center}

\setcounter{equation}{0}
\setcounter{figure}{0}
\setcounter{table}{0}
\setcounter{page}{1}
\makeatletter
\renewcommand{\theequation}{S\arabic{equation}}
\renewcommand{\thefigure}{S\arabic{figure}}
\renewcommand{\bibnumfmt}[1]{[S#1]}

\section{Linear Flavor-Wave Theory}
In this section, we give the flavor wave representation of effective spin. In the simple case that we discussed in the main text, the Hilbert space is spanned by states $|f\rangle_i\equiv|\hat{\mathbf{z}}_i\cdot\mathbf{S}_i=|f\rangle$ with $f=0,\pm 1$ for each site $i$. Then a set of SU(3) generators can be constructed as $G_f^{f'}(i)=|f\rangle_i\langle f'|_i$ with and a normalization condition $\sum_f |f\rangle_i\langle f|_i=1$.

Under this basis, the spin ladder operators can be written as
\begin{align}
    S_i^+&=\sum_{ff'}\langle f|S_i^+|f' \rangle|f \rangle\langle f'|=\sqrt{2}\left[G_1^0(i)+G_0^{\bar{1}}(i)\right]\\
    S_i^-&=\sum_{ff'}\langle f|S_i^-|f' \rangle|f \rangle\langle f'|=\sqrt{2}\left[G_{\bar{1}}^0(i)+G_0^1(i)\right].
\end{align}
Similarly,
\begin{align}
     S_i^z&=\sum_{ff'}\langle f|S_i^z|f' \rangle|f \rangle\langle f'|=G_1^1(i)-G_{\bar{1}}^{\bar{1}}(i)\\
    (S_i^z)^2&=\sum_{ff'}\langle f|(S_i^z)^2|f' \rangle|f \rangle\langle f'|=G_1^1(i)+G_{\bar{1}}^{\bar{1}}(i).
\end{align}
In the spirit of the flavor representations~\cite{joshi1999elementary,li19984}, the SU(3) algebra can be reproduced by two bosons $b$ and $\bar{b}$ as
\begin{align}
G_1^1(i)&=b^\dagger_ib_i,\\
G_{\bar{1}}^{\bar{1}}(i)&=\bar{b}^\dagger_i\bar{b}_i,\\
G_0^0(i)&=1-b^\dagger_ib_i-\bar{b}^\dagger_i\bar{b}_i,\\
G_{\bar{1}}^{1}(i)&=\bar{b}^\dagger_ib_i,\\
G_1^0(i)&=b^\dagger_i\sqrt{1-b^\dagger_ib_i-\bar{b}^\dagger_i\bar{b}_i}\approx b^\dagger_i,\\
G_{\bar{1}}^0(i)&=\bar{b}^\dagger_i\sqrt{1-b^\dagger_ib_i-\bar{b}^\dagger_i\bar{b}_i}\approx \bar{b}^\dagger_i.
\end{align}

With the above equations and $G_f^{f'}(i)={G_{f'}^{f}(i)}^\dagger$, we immediately obtain the linear-flavor wave representation Eq.~(4) in the main text.

\section{Bogoliubov-de Gennes Hamiltonian}
In this section, we derive the linear flavor-wave theory of Hamiltonian Eq.~(1), and give the explicit form of BdG Hamiltonian Eq.~(5).

The Hamiltonian Eq.~(1) can be written as
\begin{align}
    H=\sum_{\langle ij \rangle}\sum_{\alpha\beta}\mathbf{S}_i^\alpha\Lambda_{ij}^{\alpha\beta}\mathbf{S}_j^\beta+\sum_i\left[\eta(\hat{\mathbf{z}}_i\cdot\mathbf{S}_i)^2-B(\hat{\mathbf{z}}_i\cdot\mathbf{S}_i)\right],
\end{align}
where $\mathbf{S}_i^\alpha$ denotes the $\alpha$-component of $\mathbf{S}_i$ in the global coordinate, $\Lambda_{ij}$ is the coupling matrix between $\mathbf{S}_i$ and $\mathbf{S}_j$. In matrix form,
\begin{align}
    \Lambda_{ij}=\begin{pmatrix}
        J & D_{ij}^z & -D_{ij}^y\\
        -D_{ij}^z & J & D_{ij}^x\\
         D_{ij}^y & -D_{ij}^x & J
    \end{pmatrix} 
\end{align}
with $J$ the exchange coupling, $\mathbf{D}_{ij}=(D_{ij}^x,D_{ij}^y,D_{ij}^z)$ the DM interaction for bond $ij$.

Since the linear flavor-wave representation of $\mathbf{S}_i$ is defined in the local coordinate $(\hat{\mathbf{x}}_i,\hat{\mathbf{y}}_i,\hat{\mathbf{z}}_i)$ as
\begin{align}
    \left\{\begin{array}{lll}
     \hat{\mathbf{x}}_i\cdot\mathbf{S}_i=\frac{1}{\sqrt{2}}(b_i+b_i^\dagger+\bar{b}_i+\bar{b}_i^\dagger),\\
     \hat{\mathbf{y}}_i\cdot\mathbf{S}_i=\frac{i}{\sqrt{2}}(b_i-b_i^\dagger-\bar{b}_i+\bar{b}_i^\dagger),\\
     \hat{\mathbf{z}}_i\cdot\mathbf{S}_i=b^\dagger_i b_i-\bar{b}^\dagger_i \bar{b}_i,     
    \end{array}
    \right.
\end{align}
with local Ising axis $\hat{\mathbf{z}}_i=(\cos\theta_i\cos\phi_i,\cos\theta_i\sin\phi_i,\sin\theta_i)$, one needs to rotate $\mathbf{S}_i$ in Eq.~(S1) into the local coordinate by
\begin{align}
    \mathbf{S}_i=
\begin{pmatrix}
    \sin\theta_i\cos\phi_i & -\sin\phi_i & \cos\theta_i\cos\phi_i\\
    \sin\theta_i\sin\phi_i & \cos\phi_i & \cos\theta_i\sin\phi_i\\
    -\cos\theta_i & 0 & \sin\theta_i
\end{pmatrix}
    \begin{pmatrix}
        \hat{\mathbf{x}}_i\cdot\mathbf{S}_i\\
        \hat{\mathbf{y}}_i\cdot\mathbf{S}_i\\
        \hat{\mathbf{z}}_i\cdot\mathbf{S}_i
    \end{pmatrix}\equiv R_i \begin{pmatrix}
        \hat{\mathbf{x}}_i\cdot\mathbf{S}_i\\
        \hat{\mathbf{y}}_i\cdot\mathbf{S}_i\\
        \hat{\mathbf{z}}_i\cdot\mathbf{S}_i
    \end{pmatrix}
\end{align}
Correspondingly, the coupling matrix $\Lambda_{ij}$ transforms as $\tilde{\Lambda}_{ij}=R_i^T\Lambda_{ij}R_j$ so that
\begin{align}
    \mathbf{S}_i\Lambda_{ij}\mathbf{S}_j=\begin{pmatrix}
        \hat{\mathbf{x}}_i\cdot\mathbf{S}_i,\ \hat{\mathbf{y}}_i\cdot\mathbf{S}_i,\ \hat{\mathbf{z}}_i\cdot\mathbf{S}_i
    \end{pmatrix}\tilde{\Lambda}_{ij}\begin{pmatrix}
        \hat{\mathbf{x}}_j\cdot\mathbf{S}_j\\
        \hat{\mathbf{y}}_j\cdot\mathbf{S}_j\\
        \hat{\mathbf{z}}_j\cdot\mathbf{S}_j
    \end{pmatrix}.
\end{align}
To the quadratic order, we obtain
\begin{align}
    H&=\sum_{\langle ij \rangle}\frac{1}{2}\left[(\tilde{\Lambda}_{ij}^{xx}+\tilde{\Lambda}_{ij}^{yy}+i\tilde{\Lambda}_{ij}^{xy}-i\tilde{\Lambda}_{ij}^{yx})(b^\dagger_ib_j+\bar{b}_i\bar{b}_j^\dagger+b^\dagger_i\bar{b}^\dagger_j+\bar{b}_ib_j)\right.\nonumber\\
    &\left.+(\tilde{\Lambda}_{ij}^{xx}-\tilde{\Lambda}_{ij}^{yy}+i\tilde{\Lambda}_{ij}^{xy}+i\tilde{\Lambda}_{ij}^{yx})(b_ib_j+\bar{b}^\dagger_i\bar{b}_j^\dagger+b_i\bar{b}^\dagger_j+\bar{b}^\dagger_ib_j)+h.c.\right]\nonumber\\
    &+\sum_i\eta(b^\dagger_i b_i+\bar{b}^\dagger_i \bar{b}_i)-B(b^\dagger_i b_i-\bar{b}^\dagger_i \bar{b}_i)
\end{align}
With $D_{3d}$ symmetry, for the three sublattices denoted as $m=1$, $2$, $3$, we have Ising axes as
\begin{align}
\hat{\mathbf{z}}_m=\left(\sin{\frac{2\pi m}{3}}\cos\theta, -\cos{\frac{2\pi m}{3}}\cos\theta, \sin\theta\right),
\end{align}
and the DM vectors as
\begin{align}
    \left\{\begin{array}{lll}
    &\mathbf{D}_{12}=-\mathbf{D}_{21}=(0,-D_p,D_z),\\
    &\mathbf{D}_{23}=-\mathbf{D}_{32}=(\frac{\sqrt{3}}{2}D_p,\frac{1}{2}D_p,D_z),\\
    &\mathbf{D}_{31}=-\mathbf{D}_{13}=(-\frac{\sqrt{3}}{2}D_p,\frac{1}{2}D_p,D_z).
    \end{array}\right.
\end{align}
After the Fourier transform, we can obtain a BdG Hamiltonian that preserves particle-hole symmetry 
\begin{align}
    H=\frac{1}{2}\sum_\mathbf{k}\Psi_\mathbf{k}^\dagger H_\mathbf{k}\Psi_\mathbf{k}=\frac{1}{2}\sum_\mathbf{k}\Psi_\mathbf{k}^\dagger \begin{pmatrix}
        A_\mathbf{k} & B_\mathbf{k}\\
        B^*_\mathbf{-k} & A^*_\mathbf{-k}
    \end{pmatrix}\Psi_\mathbf{k}
\end{align}
with
\begin{align}
    A_\mathbf{k}=\begin{pmatrix}
        \eta-B\sin\theta & 0 & f_{3\mathbf{k}} & g_{3\mathbf{k}} & f^*_{2\mathbf{k}} & g_{2\mathbf{k}}\\
        0 & \eta+B\sin\theta & g_{3\mathbf{k}} & f^*_{3\mathbf{k}} & g_{2\mathbf{k}} & f_{2\mathbf{k}}\\
        f^*_{3\mathbf{k}} & g_{3\mathbf{k}} & \eta-B\sin\theta & 0 & f_{1\mathbf{k}} & g_{1\mathbf{k}}\\
        g_{3\mathbf{k}} & f_{3\mathbf{k}} & 0 & \eta+B\sin\theta & g_{1\mathbf{k}} & f^*_{1\mathbf{k}}\\
        f_{2\mathbf{k}} & g_{2\mathbf{k}} & f^*_{1\mathbf{k}} & g_{1\mathbf{k}} & \eta-B\sin\theta & 0\\
        g_{2\mathbf{k}} & f^*_{2\mathbf{k}} & g_{1\mathbf{k}} & f_{1\mathbf{k}} & 0 & \eta+B\sin\theta\\
    \end{pmatrix}
\end{align}
and
\begin{align}
    B_\mathbf{k}=\begin{pmatrix}
        0 & 0 & g_{3\mathbf{k}} & f_{3\mathbf{k}} & g_{2\mathbf{k}} & f^*_{2\mathbf{k}}\\
        0 & 0 & f^*_{3\mathbf{k}} & g_{3\mathbf{k}} & f_{2\mathbf{k}} & g_{2\mathbf{k}}\\
        g_{3\mathbf{k}} & f^*_{3\mathbf{k}} & 0 & 0 & g_{1\mathbf{k}} & f_{1\mathbf{k}}\\
        f_{3\mathbf{k}} & g_{3\mathbf{k}} & 0 & 0 & f^*_{1\mathbf{k}} & g_{1\mathbf{k}}\\
        g_{2\mathbf{k}} & f_{2\mathbf{k}} & g_{1\mathbf{k}} & f^*_{1\mathbf{k}} & 0 & 0\\
        f^*_{2\mathbf{k}} & g_{2\mathbf{k}} & f_{1\mathbf{k}} & g_{1\mathbf{k}} & 0 & 0\\
    \end{pmatrix},
\end{align}
where $f_{m\mathbf{k}}=\left[\frac{1}{2}J(1-3\sin^2\theta)+\frac{\sqrt{3}}{2}D_z(1+\sin^2\theta)-\sqrt{3}D_p\sin\theta\cos\theta-i(\sqrt{3}J\sin\theta+D_z\sin\theta+D_p\cos\theta)\right]\cos\left(\mathbf{k}\cdot\boldsymbol{\delta}_m\right)$ and $g_{m\mathbf{k}}=\left[\frac{3}{2}(J-\frac{1}{\sqrt{3}}D_z)\cos^2\theta-\sqrt{3}D_p\cos\theta\sin\theta\right]\cos\left(\mathbf{k}\cdot\boldsymbol{\delta}_m\right)$.

\begin{figure}
\includegraphics[width=0.5\textwidth]{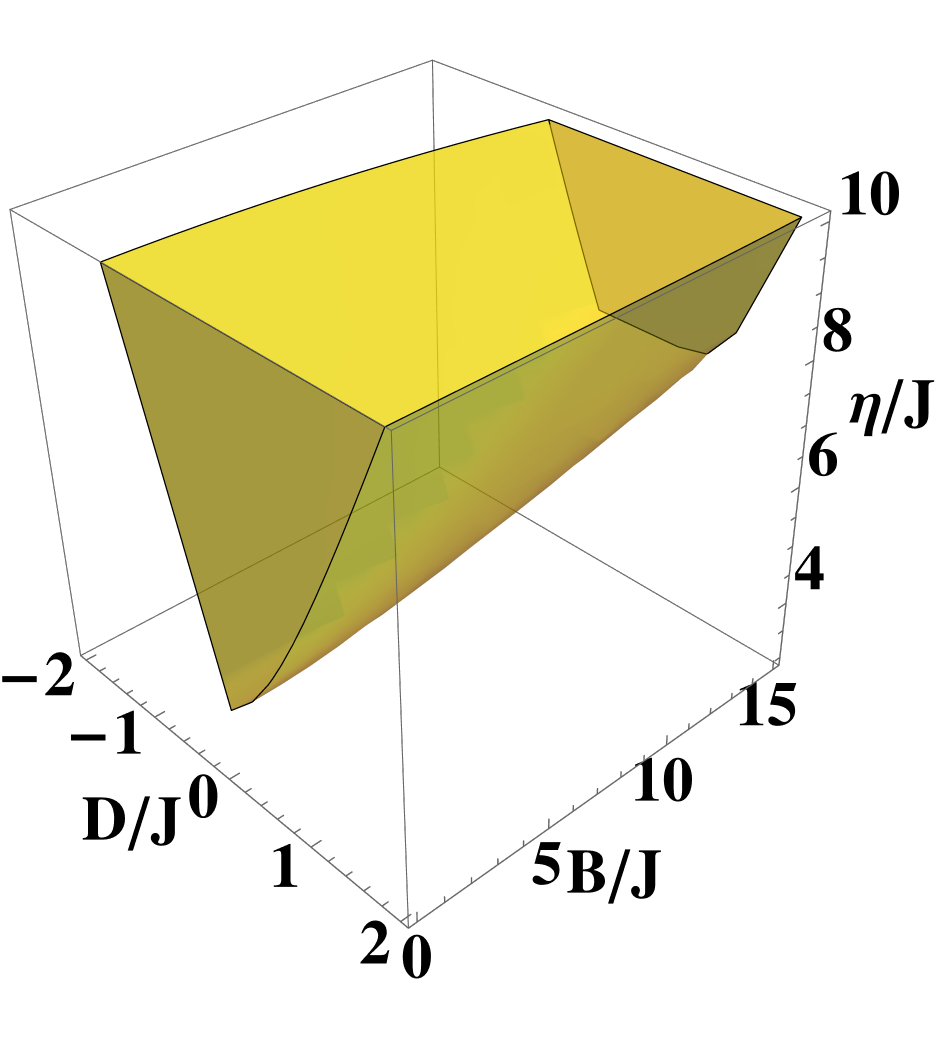}
\caption{\label{fig:Phase} (Color online.) The quantum paramagnetic phase (shown as yellow region) with $\sin\theta=1/3$ and $\sqrt{3}D_p=\sqrt{3/2}D_z=D$.}
\end{figure}

Since the commutator gives $[\Psi_\mathbf{k},\Psi^\dagger_\mathbf{k}]=\begin{pmatrix}
    1 & 0\\
    0 & -1
\end{pmatrix}\otimes I_{3\times 3}\equiv \Sigma_z$, if we perform a Bogoliubov transformation $\Psi_\mathbf{k}=T_\mathbf{k}\psi_\mathbf{k}$ to diagonalize $H_\mathbf{k}$ while preserve the commutator, i.e. $[\psi_\mathbf{k},\psi^\dagger_\mathbf{k}]=\Sigma_z$, then
\begin{align}
    \Sigma_z=[\Psi_\mathbf{k},\Psi^\dagger_\mathbf{k}]=T_\mathbf{k}[\psi_\mathbf{k},\psi^\dagger_\mathbf{k}]T^\dagger_\mathbf{k}=T_\mathbf{k}\Sigma_zT^\dagger_\mathbf{k},\quad\text{and}\quad T^\dagger_\mathbf{k}H_\mathbf{k}T_\mathbf{k}=\begin{pmatrix}
        E_\mathbf{k} & 0\\
        0 & E_\mathbf{-k}
    \end{pmatrix},
\end{align}
where $E_\mathbf{k}$ is a $6\times 6$ diagonal matrix with elements the eigen-energy and $E_{-\mathbf{k}}$ is the particle-hole symmetric partner of $E_\mathbf{k}$. As we mentioned in the main text, for any $E_\mathbf{k}$, the positiveness of the diagonal elements determines the mean-field phase diagram of this quantum paramagnetic phase, and we show the diagram in Fig.~\ref{fig:Phase} with parameters $\sin\theta=1/3$, $\sqrt{3}D_p=\sqrt{3/2}D_z=D$ as an example.

\begin{figure}
\includegraphics[width=0.96\textwidth]{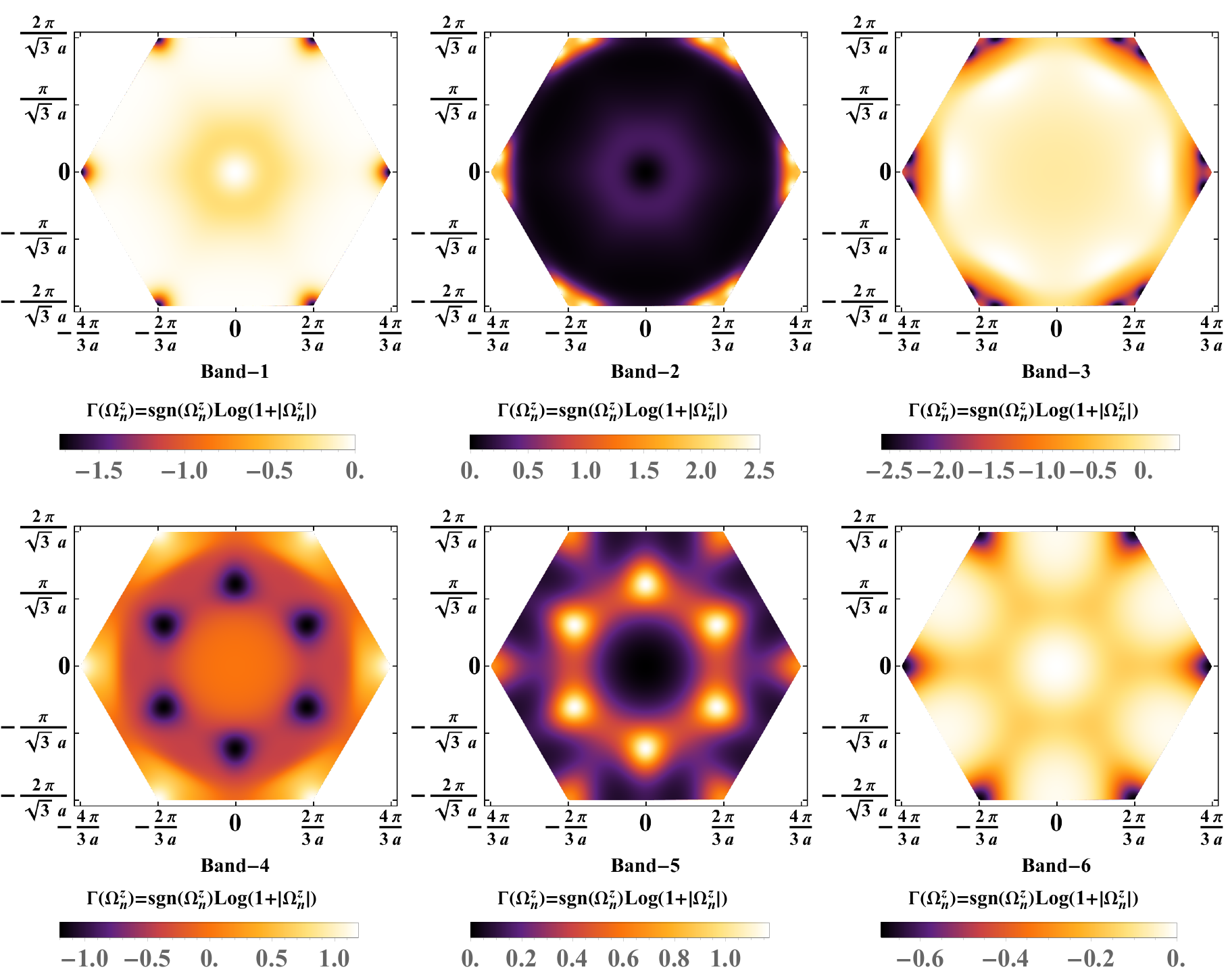}
\caption{\label{fig:BC} (Color online.) The distribution of Berry curvature (in log scale) in the momentum space from the lowest band (band-1) to the highest band (band-6) with a parameter choice as  $\sin\theta=1/3$, $\eta/J=7.0$, $\sqrt{3}D_p=\sqrt{3/2}D_z=D$, and $B/J=0.5$. }
\end{figure}

\begin{figure}
\includegraphics[width=0.6\textwidth]{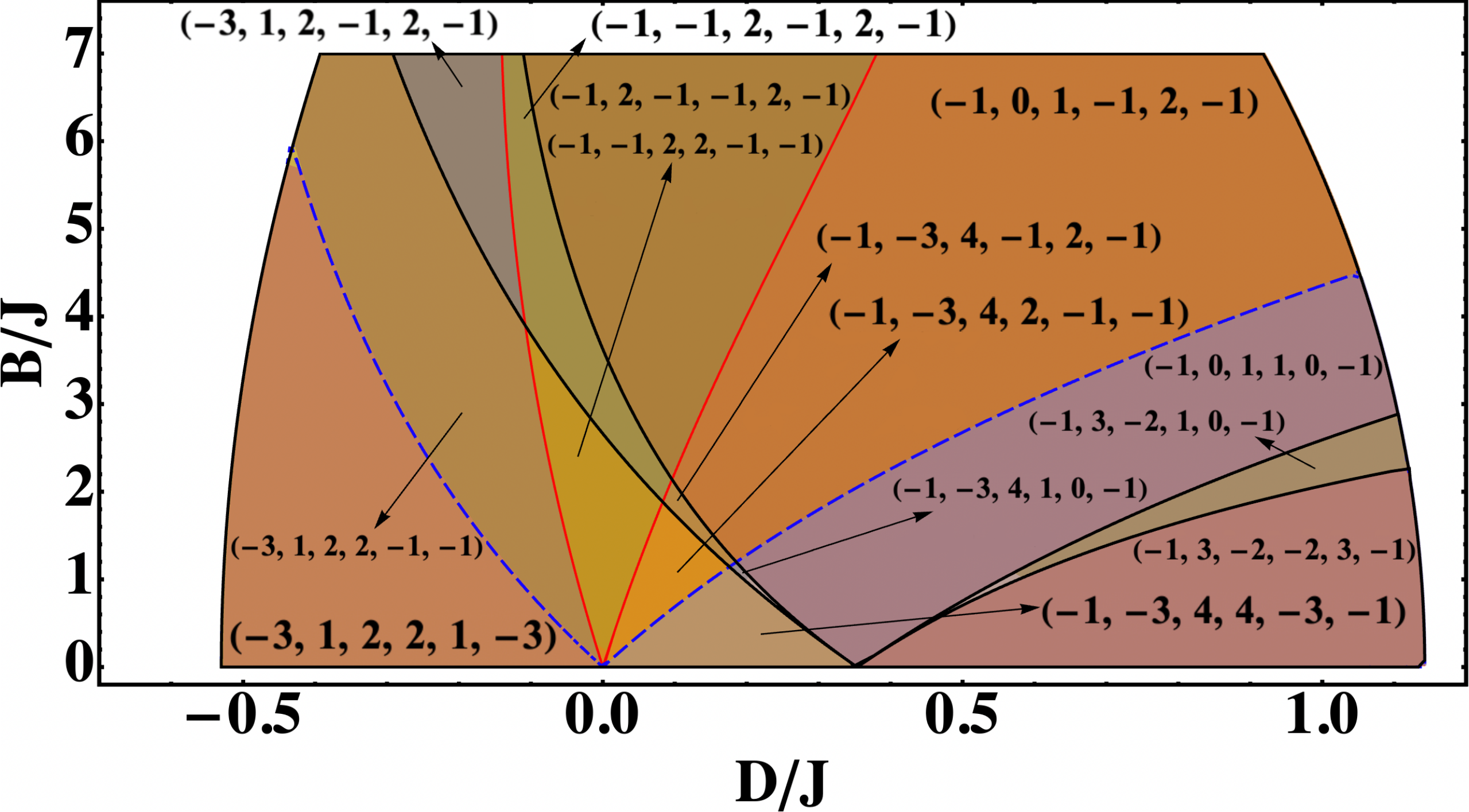}
\caption{\label{fig:CN_full} Diagram of all six band Chern numbers distributions. The Chern numbers are listed from bottom to top. The parameters are the same as Fig.~3 in the main text. In addition to the band touching at $\boldsymbol{\Gamma}$ and $\mathbf{M}$ denoted by the red solid and black thick lines. There is band-touching at $\mathbf{K}$ denoted by blue dashed lines that give rise to more complicated topological structures here.}
\end{figure}

It can also be checked that $i\frac{d}{dt}\psi_{\mathbf{k}}=[\psi_\mathbf{k},H]=\Sigma_z H_{\mathbf{k}}\psi_{\mathbf{k}}$, and thus a proper Lagrangian should be
\begin{align}
    \mathcal{L}_\mathbf{k}=i\frac{d}{dt}-\Sigma_z H_\mathbf{k}.
\end{align}
Therefore, a bosonic vector potential $\boldsymbol{\mathcal{A}}_{n\mathbf{k}}$ and Berry curvature $\boldsymbol{\Omega}_{n\mathbf{k}}$ for $\psi_{\mathbf{k}}$ can be defined as
\begin{align}
    \boldsymbol{\mathcal{A}}_{n\mathbf{k}}=i\langle \psi_{n\mathbf{k}}|\Sigma_z\boldsymbol{\nabla}_\mathbf{k}|\psi_{n\mathbf{k}}\rangle,\text{ and }\boldsymbol{\Omega}_{n\mathbf{k}}=\boldsymbol{\nabla}_{\mathbf{k}}\times\boldsymbol{\mathcal{A}}_{n\mathbf{k}}.
\end{align}
With the expression above, we show in Fig.~\ref{fig:BC} the Berry curvature distribution $\Omega_{n\mathbf{k}}^z$ in the Brillouin zone. These finite values of Berry curvature lead to the non-zero bosonic band Chern numbers that we discussed in the main text. We also show the full diagram for all the six band Chern numbers in Fig.~\ref{fig:CN_full}.

\section{Collinear case}
In this section, we derive Eq.~(8) in the main text. We first take $\theta=\pi/2$ and $D_p=0$ into Eq.~(S9)-(S11), and to further simplify the expression, we then perform a gauge transformation as $b_{m\mathbf{k}}\rightarrow ie^{-i\frac{2\pi m}{3}}b_{m\mathbf{k}}$ and $\bar{b}_{m\mathbf{k}}\rightarrow -ie^{i\frac{2\pi m}{3}}\bar{b}_{m\mathbf{k}}$. The $12\times 12$ Hamiltonian matrix in the basis 
 $\Psi_\mathbf{k}=\left(b_{1\mathbf{k}},\bar{b}_{1\mathbf{k}},...,b_{3\mathbf{k}},\bar{b}_{3\mathbf{k}},b^\dagger_{1,-\mathbf{k}},\bar{b}^\dagger_{1,-\mathbf{k}},...,b^\dagger_{3,-\mathbf{k}},\bar{b}^\dagger_{3,-\mathbf{k}}\right)^T$ is expressed as
\begin{align}
\text{${\tiny H_\mathbf{k}=2\left(\begin{array}{cccccccccccc}
 \frac{\eta -B}{2} & 0 & \tilde{J}  \cos\mathbf{k}_3 & 0 & \tilde{J}  ^*\cos\mathbf{k}_2 & 0 & 0 & 0 & 0 & \tilde{J}  \cos\mathbf{k}_3 & 0 & \tilde{J}  ^*\cos\mathbf{k}_2\\
 0 & \frac{\eta +B}{2} & 0 & \tilde{J}  ^* \cos\mathbf{k}_3 & 0 & \tilde{J}   \cos\mathbf{k}_2 & 0 & 0 & \tilde{J}  ^* \cos\mathbf{k}_3 & 0 & \tilde{J}   \cos\mathbf{k}_2 & 0 \\
 \tilde{J}  ^* \cos\mathbf{k}_3 & 0 & \frac{\eta -B}{2} & 0 & \tilde{J}   \cos\mathbf{k}_1 & 0 & 0 & \tilde{J}  ^* \cos\mathbf{k}_3 & 0 & 0 & 0 & \tilde{J}   \cos\mathbf{k}_1 \\
 0 & \tilde{J}   \cos\mathbf{k}_3 & 0 & \frac{\eta +B}{2} & 0 & \tilde{J}  ^* \cos\mathbf{k}_1 & \tilde{J}   \cos\mathbf{k}_3 & 0 & 0 & 0 & \tilde{J}  ^* \cos\mathbf{k}_1 & 0 \\
 \tilde{J}   \cos\mathbf{k}_2 & 0 & \tilde{J}  ^* \cos\mathbf{k}_1 & 0 & \frac{\eta -B}{2} & 0 & 0 & \tilde{J}   \cos\mathbf{k}_2 & 0 & \tilde{J}  ^* \cos\mathbf{k}_1 & 0 & 0 \\
 0 & \tilde{J}  ^* \cos\mathbf{k}_2 & 0 & \tilde{J}   \cos\mathbf{k}_1 & 0 & \frac{\eta +B}{2} & \tilde{J}  ^* \cos\mathbf{k}_2 & 0 & \tilde{J}   \cos\mathbf{k}_1 & 0 & 0 & 0 \\
 0 & 0 & 0 & \tilde{J}  ^* \cos\mathbf{k}_3 & 0 & \tilde{J}   \cos\mathbf{k}_2 & \frac{\eta -B}{2} & 0 & \tilde{J}  ^* \cos\mathbf{k}_3 & 0 & \tilde{J}   \cos\mathbf{k}_2 & 0 \\
 0 & 0 & \tilde{J}   \cos\mathbf{k}_3 & 0 & \tilde{J}  ^* \cos\mathbf{k}_2 & 0 & 0 & \frac{\eta +B}{2} & 0 & \tilde{J}   \cos\mathbf{k}_3 & 0 & \tilde{J}  ^* \cos\mathbf{k}_2 \\
 0 & \tilde{J}   \cos\mathbf{k}_3 & 0 & 0 & 0 & \tilde{J}  ^* \cos\mathbf{k}_1 & \tilde{J}   \cos\mathbf{k}_3 & 0 & \frac{\eta -B}{2} & 0 & \tilde{J}  ^* \cos\mathbf{k}_1 & 0 \\
 \tilde{J}  ^* \cos\mathbf{k}_3 & 0 & 0 & 0 & \tilde{J}   \cos\mathbf{k}_1 & 0 & 0 & \tilde{J}  ^* \cos\mathbf{k}_3 & 0 & \frac{\eta +B}{2} & 0 & \tilde{J}   \cos\mathbf{k}_1 \\
 0 & \tilde{J}  ^* \cos\mathbf{k}_2 & 0 & \tilde{J}   \cos\mathbf{k}_1 & 0 & 0 & \tilde{J}  ^* \cos\mathbf{k}_2 & 0 & \tilde{J}   \cos\mathbf{k}_1 & 0 & \frac{\eta -B}{2} & 0 \\
 \tilde{J}   \cos\mathbf{k}_2 & 0 & \tilde{J}  ^* \cos\mathbf{k}_1 & 0 & 0 & 0 & 0 & \tilde{J}   \cos\mathbf{k}_2 & 0 & \tilde{J}  ^* \cos\mathbf{k}_1 & 0 & \frac{\eta +B}{2}
\end{array}\right)}$}.
\end{align}
Now with the basis transformation that we mentioned in the main text: $u_{m\mathbf{k}}=\frac{1}{\sqrt{2}}(b_{m\mathbf{k}}+\bar{b}_{m,-\mathbf{k}}^\dagger)$ and $p_{m\mathbf{k}}=\frac{i}{\sqrt{2}} (\bar{b}_{m,-\mathbf{k}}^\dagger-b_{m\mathbf{k}})$, we can obtain $H=\frac{1}{2}\sum_\mathbf{k}\Phi^\dagger_\mathbf{k}\tilde{H}_\mathbf{k}\Phi_\mathbf{k}$ with $\Phi_\mathbf{k}=\left(u_{1\mathbf{k}},...,u_{3\mathbf{k}},p_{1\mathbf{k}},...,p_{3\mathbf{k}},u^\dagger_{1,-\mathbf{k}},...,u^\dagger_{3,-\mathbf{k}},p^\dagger_{1,-\mathbf{k}},...,p^\dagger_{3,-\mathbf{k}}\right)^T$, where
\begin{align}
    \tilde{H}_\mathbf{k}=\begin{pmatrix}
        2M_\mathbf{k} & -iB I_3 & 0 & 0\\
        iB I_3 & \eta I_3 & 0 &0\\
        0 & 0 & 2M^*_\mathbf{k} & iB I_3\\
        0 & 0 & -iB I_3 & \eta I_3\\
    \end{pmatrix},
\end{align}
or alternatively,
\begin{align}
    H&=\frac{1}{2\eta^{-1}}\sum_\mathbf{k}(\mathbf{p}_\mathbf{k}^\dagger-i\frac{B}{\eta}\mathbf{u}^\dagger_\mathbf{k})(\mathbf{p}_\mathbf{k}+i\frac{B}{\eta}\mathbf{u}_\mathbf{k})+\frac{1}{2}\sum_\mathbf{k}\mathbf{u}_\mathbf{k}^\dagger(2M_\mathbf{k}-\frac{B^2}{\eta})\mathbf{u}_\mathbf{k}\nonumber\\
    &+\frac{1}{2\eta^{-1}}\sum_\mathbf{k}(\mathbf{p}_{-\mathbf{k}}+i\frac{B}{\eta}\mathbf{u}_{-\mathbf{k}})(\mathbf{p}_{-\mathbf{k}}^\dagger-i\frac{B}{\eta}\mathbf{u}^\dagger_{-\mathbf{k}})+\frac{1}{2}\sum_\mathbf{k}\mathbf{u}_{-\mathbf{k}}(2M^*_\mathbf{k}-\frac{B^2}{\eta})\mathbf{u}_{-\mathbf{k}}^\dagger\nonumber\\
    &=\frac{1}{2(2\eta)^{-1}}\sum_\mathbf{k}(\mathbf{p}_\mathbf{k}^\dagger-i\frac{B}{\eta}\mathbf{u}^\dagger_\mathbf{k})(\mathbf{p}_\mathbf{k}+i\frac{B}{\eta}\mathbf{u}_\mathbf{k})+\frac{1}{2}\sum_\mathbf{k}\mathbf{u}_\mathbf{k}^\dagger(4M_\mathbf{k}-\frac{2B^2}{\eta})\mathbf{u}_\mathbf{k}-B
\end{align}
as written in Eq.~(8) of the main text (up to a constant), where we have used the fact that $M_\mathbf{k}=M_{-\mathbf{k}}=M_\mathbf{k}^\dagger$ and $[u^\dagger_{m\mathbf{k}},p_{m'\mathbf{k}'}]=i\delta_{mm'}\delta_{\mathbf{k}\mathbf{k}'}$.

Since $M_\mathbf{k}$ is Hermitian, we can diagonalize $2\eta M_\mathbf{k}$ by a unitary matrix $Q_\mathbf{k}$ as $2\eta Q^\dagger_\mathbf{k}M_\mathbf{k}Q_\mathbf{k}=\text{Diag}(\tilde{E}^2_{1\mathbf{k}},\tilde{E}^2_{2\mathbf{k}},\tilde{E}^2_{3\mathbf{k}})\equiv\tilde{E}^2_\mathbf{k}$ with $Q_\mathbf{k}^\dagger Q_\mathbf{k}=I_3$.
Then, it can be found that 
\begin{align}
    \tilde{Q}_\mathbf{k}=\begin{pmatrix}
        Q_\mathbf{k}\sqrt{\frac{\eta}{2\tilde{E}_\mathbf{k}}} & & & Q_\mathbf{k}\sqrt{\frac{\eta}{2\tilde{E}_\mathbf{k}}}\\
          & Q^*_\mathbf{k}\sqrt{\frac{\eta}{2\tilde{E}_\mathbf{k}}} & Q^*_\mathbf{k}\sqrt{\frac{\eta}{2\tilde{E}_\mathbf{k}}} &\\
         & -iQ^*_\mathbf{k}\sqrt{\frac{\tilde{E}_\mathbf{k}}{2\eta}} & iQ^*_\mathbf{k}\sqrt{\frac{\tilde{E}_\mathbf{k}}{2\eta}} & \\
          -iQ_\mathbf{k}\sqrt{\frac{\tilde{E}_\mathbf{k}}{2\eta}} & & & iQ_\mathbf{k}\sqrt{\frac{\tilde{E}_\mathbf{k}}{2\eta}}
    \end{pmatrix}
\end{align}
can diagonalize $\tilde{H}_\mathbf{k}$ as
\begin{align}
    \tilde{Q}_\mathbf{k}^\dagger \tilde{H}_\mathbf{k}\tilde{Q}_\mathbf{k}=\begin{pmatrix}
        \tilde{E}_\mathbf{k}- B & & &\\
         & \tilde{E}_\mathbf{k}+ B & &\\
         & &\tilde{E}_{-\mathbf{k}}- B &\\
         & & &\tilde{E}_{-\mathbf{k}}+ B\\
    \end{pmatrix},
\end{align}
while transforming the commutator $[\Phi_\mathbf{k},\Phi^\dagger_\mathbf{k}]=I_2\otimes\begin{pmatrix}0 & i\\-i & 0\end{pmatrix}\otimes I_3$ into $\tilde{Q}_\mathbf{k}^\dagger[\Phi_\mathbf{k},\Phi^\dagger_\mathbf{k}]\tilde{Q}_\mathbf{k}=\Sigma_z=\begin{pmatrix}1 & 0\\0 & -1\end{pmatrix}\otimes I_6$ the canonical bosonic commutator with particle (hole) eigen-energy $E_{\pm\mathbf{k}}=\begin{pmatrix}
    \tilde{E}_{\pm\mathbf{k}}- B &\\
         & \tilde{E}_{\pm\mathbf{k}}+ B
\end{pmatrix}$, and thus $\tilde{Q}_\mathbf{k}$ is the proper wavefunctions for the ``phononic'' Hamiltonian Eq.~(S17). It can be noticed that $\tilde{Q}_\mathbf{k}$ fully depends on $M_\mathbf{k}$ and does not change with non-zero $B$. Therefore, we have the conclusion in the main text that the topological properties of Hamiltonian $H$ are fully determined by $M_\mathbf{k}$.
\end{document}